\documentclass[prl,showpacs]{revtex4}
\def\comment#1{}

\usepackage{graphicx}
\begin{document}
\title{Analytic Solutions of the Ultra-relativistic Thomas-Fermi Equation}
\author{Michael Rotondo$^{2}$, Remo Ruffini$^{1,2,3}$ and She-Sheng Xue$^{1,2}$}
\email{ruffini@icra.it}
\affiliation{
$^{1}$%
ICRANet, Piazzale della Repubblica 10, 65122, Pescara, Italy \\
$^{2}$%
Department of Physics and ICRA, University of Rome `Sapienza', Piazzale A.Moro 5,  00185, Rome, Italy\\
$^{3}$%
ICRANet, University of Nice-Sophia Antipolis, 28 avenue de Valrose, 06103 Nice Cedex 2, France}
\received{ \today}

\begin{abstract}
It is well known that the ultra-relativistic Thomas-Fermi equation, amply adopted in the study of heavy nuclei, admits an exact solution for a constant proton distribution within a spherical core of radius $R_c$. Here exact solutions of a generalized ultra-relativistic Thomas-Fermi equation are presented, assuming a Wood-Saxon-like proton distribution and its further generalizations. These solutions present an overcritical electric field close to their surface. The variation of the electric fields as a function of the generalized Wood-Saxon parameters are studied.  
\end{abstract}

\maketitle

\section{Introduction}
To study the electrodynamic properties of the bulk matter at nuclear densities a step  proton distribution has been chosen \cite{rrx200613, prrx2009}. Using the Migdal et. al.  approximation, \cite{migdal7614}, the ultra-relativistic Thomas-Fermi model, which governs this problem, reads  
\begin{equation}
\frac{d^2\phi(x)}{dx^2}=\phi(x)^3-\theta(-x),
\label{eqless5g}
\end{equation}
where the proton density, $n_p$, and the Coulomb potential at the center, $V(0)$, are given by
\begin{equation}
x=k(r-R_c),
\label{eqless5gA}
\end{equation}
\begin{equation}
k=2 \sqrt{\alpha}(\pi/6)^{1/6}n_p^{1/3},
\label{eq1}
\end{equation}
\begin{equation}
eV(0)=( 3 \pi^2 n_p)^{1/3}.
\label{eqless5gB}
\end{equation}
The equation (\ref{eqless5g}) admits the exact solution 
\begin{equation}
 \phi(x) = \left\{\begin{array}{ll} 1-3\left[1+2^{-1/2}\sinh(a-\sqrt{3}x)\right]^{-1}, &  \quad x<0, \\
\frac {\sqrt{2}}{(x+b)}, & \quad x>0,
\end{array}\right.
\label{popovs2}
\end{equation}
where integration constants $a$ and $b$ are: $\sinh a=11\sqrt{2}$, $a=3.439$; $b=(4/3)\sqrt{2}$.\cite{migdal7614}.

\section{Generalized Ultra-relativistic Thomas-Fermi Equation}
In this section we want to look for exact solutions to a generalized ultra-relativistic Thomas-Fermi equation
\begin{equation}
\frac{d^2\phi(x)}{dx^2}=\phi(x)^3-f_p\theta(-x),
\label{eqless5g1}
\end{equation}
where  
\begin{equation}
  \left\{\begin{array}{ll} f_{p}(x_b)\stackrel{}{\rightarrow}0, & 0 \leq x_b \leq \infty \\
f_{p}(-\infty)\stackrel{}{\rightarrow}1,\\ f'_{p}(x)\leq 0,& {\rm for}\quad all \quad x
\end{array}\right.
\label{popovsab}
\end{equation}

It is possible to write several distinct infinite $b$-dependent sets of analytic solutions to the Thomas-Fermi  Eq. $( \ref{eqless5g1})$.\\
\vskip0.2cm
\noindent{\it - Set 1}\hskip0.3cm 
\begin{equation}
 \phi(x;b)= \left\{\begin{array}{ll} \frac{1}{2}-\frac{1}{\pi}\arctan(bx) , &  {\rm for}\quad x<x_{b}, \\
\frac{\alpha}{\beta+x}, & {\rm for}\quad x>x_{b},
\end{array}\right.
\label{popovsb}
\end{equation}
and leads to the following set of proton profiles 
\begin{equation}
 f_{p}(x;b)= \left\{\begin{array}{ll} \frac{1}{\pi^{3}}\left((\frac{\pi}{2}-\arctan(bx)\right)^{3} -\frac{2b^{3}x}{\pi(1+b^{2}x^{2})^{2}}, &  {\rm for}\quad x<x_{b}, \\
0, & {\rm for}\quad x>x_{b},
\end{array}\right.
\label{popovsa}
\end{equation}
where $\alpha$, $\beta$ are real constants given by
\begin{equation}
 \left\{\begin{array}{ll} \alpha=\frac{\pi(1+b^{2}x_{b})}{b}(\phi(x_{b};b))^{2} ,  \\
\beta=\frac{\pi(1+b^{2}x_{b})}{b}\phi(x_{b};b)-x_{b},
\end{array}\right.
\label{popovsc}
\end{equation}
because of the continuity of $\phi(x;b)$, $\phi'(x;b)$ in $x_{b}$.\\
The electric field for $x<x_{b}$, is given by
\begin{equation}
E(x;b)=\frac{2}{(3\pi)^{1/2}}e^2V(0)^{2}\frac{1}{\pi}\frac{b}{1+(bx)^{2}}.
\label{eq1a}
\end{equation}
The parameter $b$ describes the width $2\delta$ (in $cm$) of the transition layer near the edge of the core
\begin{equation}
b=\frac{1}{k\delta}.
\label{w-sa}
\end{equation}
Precisely $2\delta$ is the width of the transition layer of the core in which the electric field goes from its maximum to the half of its maximum.
Now, let $b_{c}$ be the value of $b$ such that the electric field $E(x=0;b_{c})$ is equal to the critical field  $E_{c}$. Then
\begin{equation}
b_{c;Set1}\approx  \frac{1}{0.8}\frac{E_{c}}{E_{max}},
\label{eq1A}
\end{equation}
and
\begin{equation}
\delta_{c;Set1}=\left[\frac{1}{2^{29/6}} \frac{27}{5}\right]\left[\frac{\hbar}{mc}\right]a_{0}n_{p}^{1/3}(cm).
\label{eq1B}
\end{equation}
where $E_{max}$ is the electric field at $x=0$ to the step-proton distribution.
We see that  $\delta_{c}$ can be of the order of the Bohr radius $a_0$ i.e. of order of $10^{3}$ electron Compton length.


\begin{figure}[th] 
\begin{center}
\includegraphics[width=\hsize,clip]{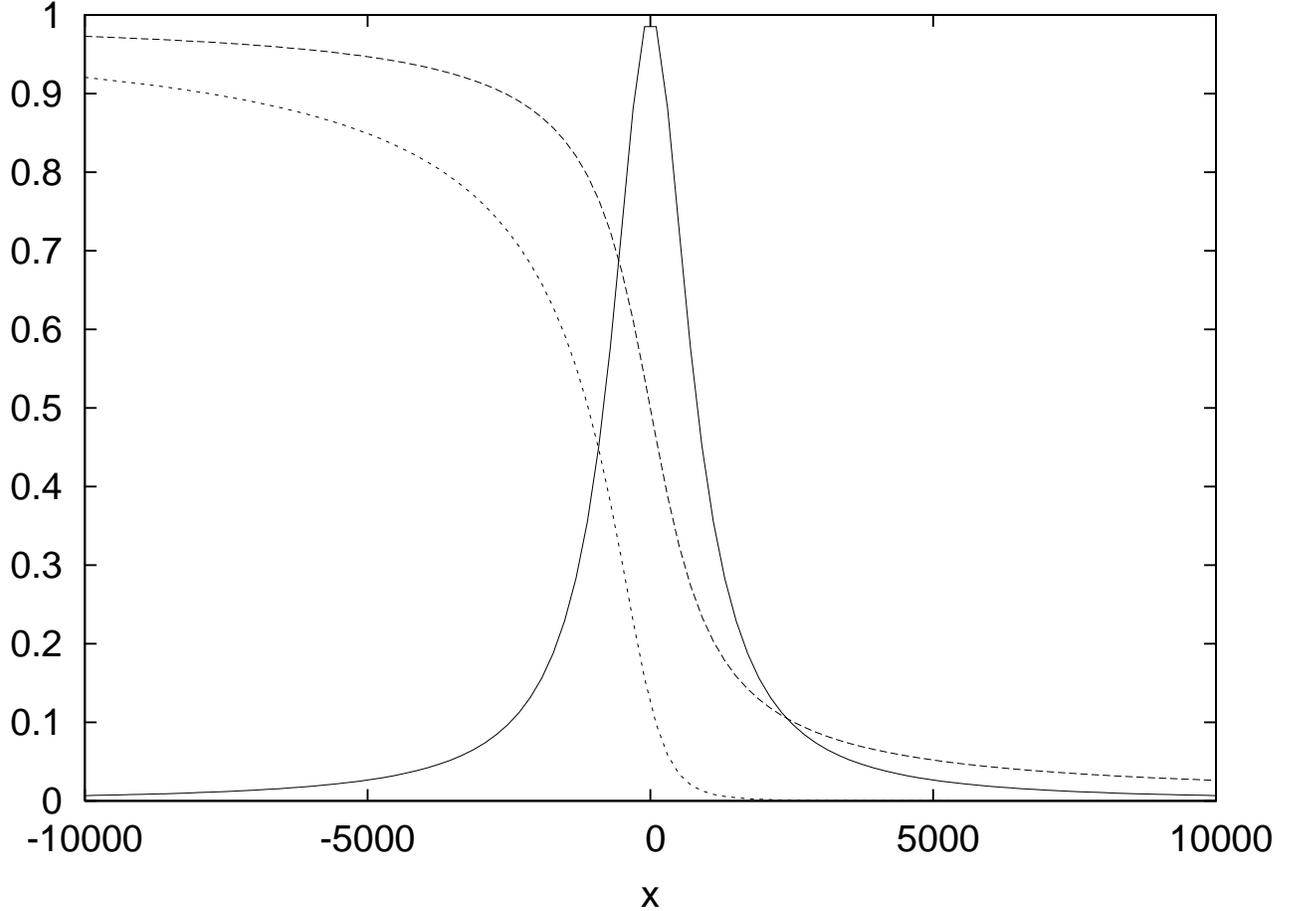}
\end{center}
\caption{The variation of the potential, the proton density and the electric field strength near the edge of the core of the $Set$ $1$ are plotted as  functions of $x$, for a proton number density at the center $n_p^0=0.57\cdot10^{37} cm^{-3}$. The dashed curve represents the function $\phi(x;b_{c})$; the dotted curve represents the distribution $f_{p}(x;b_{c})$; the solid curve represents the ratio $E(x;b_{c})/E_{c}$. }%
\label{phifE}%
\end{figure}

\vskip0.2cm
\noindent{\it - Set 2}\hskip0.3cm 
\begin{equation}
\phi(x;b)= \left\{\begin{array}{ll} \frac{1}{2}-\frac{1}{2}\tanh(bx) , &  {\rm for}\quad x<x_{b}, \\
\frac{\alpha}{\beta+x}, & {\rm for}\quad x>x_{b},
\end{array}\right.
\label{popovsb1}
\end{equation}
and leads to the following set of proton profiles
\begin{equation}
 f_{p}(x;b)= \left\{\begin{array}{ll} \left(\frac{1}{8}(1-\tanh(bx)\right)^{3} -b^{2}\tanh(bx)(1-(\tanh(bx))^{2}), &  {\rm for}\quad x<x_{b}, \\
0, & {\rm for}\quad x>x_{b},
\end{array}\right.
\label{popovsa1}
\end{equation}
where $\alpha$, $\beta$ are real constants given by
\begin{equation}
 \left\{\begin{array}{ll} \alpha=\frac{1}{b^{2}}(\frac{1}{2}-\frac{\tanh(bx_{b})}{1+(\tanh(bx_{b}))^2}) ,  \\
\beta=(\frac{\tanh(bx_{b})}{b^{2}-1-(\tanh(bx_{b}))^2})-x_{b},
\end{array}\right.
\label{popovsc1}
\end{equation}
because of the continuity of $\phi(x;b)$, $\phi'(x;b)$ in $x_{b}$.\\
The electric field for $x<x_{b}$, is given by
\begin{equation}
E(x;b)=\frac{2}{(3\pi)^{1/2}}e^2V(0)^{2}\frac{1}{2}\frac{b}{cosh^{2}(bx)}.
\label{eq1C}
\end{equation}
The parameter $b$ as above, describes the width $2\delta$ (in $cm$) of the transition layer near the edge of the core.
Precisely 
\begin{equation}
b_{SET2}=b_{SET1}\frac{ln(2\sqrt(2)+3)}{2}.
\label{w-saB}
\end{equation}
Now, let $b_{c}$ be the value of $b$ such that the electric field $E(x=0;b_{c})$ is equal to the critical field  $E_{c}$. Then
\begin{equation}
b_{c;Set2}\approx  \frac{1}{1.3}\frac{E_{c}}{E_{max}},
\label{eq1D}
\end{equation}
and
\begin{equation}
\delta_{c;Set2}=\frac{ln(2\sqrt(2)+3)}{2}\delta_{c;Set1}(cm).
\label{eq1S}
\end{equation}
We note that 
\begin{equation}
\frac{1}{2}-\frac{1}{2}\tanh(bx)=\frac{1}{1+e^{(2bx)}}
\label{w-s}
\end{equation}
which is well known in nuclear physics as Wood-Saxon profile.\\
The Wood-Saxon profile can be generalized by
\begin{equation}
\phi(x;a,b)= \left\{\begin{array}{ll} 1-\frac{1}{\left[1+ae^{(-bx)}\right]^{1/a}} , &  {\rm for}\quad x<x_{b}, \\
\frac{\alpha}{\beta+x}, & {\rm for}\quad x>x_{b},
\end{array}\right.
\label{popovsb1A}
\end{equation}
with the following set of proton profiles
\begin{equation}
 f_{p}(x;a,b)= \left\{\begin{array}{ll} \left\{1-\frac{1}{\left[1+ae^{(-bx)}\right]^{1/3}}\right\}^{3}+\frac{b^2e^{-bx}(e^{-bx}-1)}{(1+ae^{-bx})^{1/a}(1+ae^{-bx})^{2}}, & {\rm for}\quad x>x_{b}\\
0, & {\rm for}\quad x>x_{b},
\end{array}\right.
\label{popovsa1A}
\end{equation}
where $\alpha$, $\beta$ are real constants given by
\begin{equation}
 \left\{\begin{array}{ll} \alpha=\left[1-\frac{1}{(1+ae^{-bx_{b}})^{1/a}}\right]\frac{\left[(1+ae^{-bx_{b}})^{1/a}-1\right]\left[1+ae^{-bx_b}\right]}{be^{-bx_{b}}},  \\
\beta=-x_{b}+\frac{\left[(1+ae^{-bx_{b}})^{1/a}-1\right]\left[1+ae^{-bx_b}\right]}{be^{-bx_{b}}}.
\end{array}\right.
\label{popovsc1A}
\end{equation}
because of the continuity of $\phi(x;a,b)$, $\phi'(x;a,b)$ in $x_{b}$.\\
We have
\begin{eqnarray}
\phi'(x;a,b)=-\frac{be^{-bx}}{(1+ae^{-bx})^{1/a}(1+ae^{-bx})},
\label{dless1}
\end{eqnarray}
hence the maximum of $E(x;a,b)$ is
\begin{equation}
E(x=0;a,b)= \frac{b}{(1+a)^{1/a}(1+a)}E_{max}.
\label{eq1E}
\end{equation}
\vskip0.2cm
\noindent{\it - Set 3}\hskip0.3cm 
\begin{equation}
 \phi(x;b)= \left\{\begin{array}{ll} \frac{1}{2}-\frac{1}{72}\sinh^{-1}(bx) , &  {\rm for}\quad x<x_{b}, \\
\frac{\alpha}{\beta+x}, & {\rm for}\quad x>x_{b},
\end{array}\right.
\label{popovsbAA}
\end{equation}
and leads to the following set of proton profiles
\begin{equation}
 f_{p}(x;b)= \left\{\begin{array}{ll} \left((\frac{1}{2}-\frac{1}{72}\sinh^{-1}(bx)\right)^{3} -\frac{b^{3}x}{72(1+b^{2}x^{2})^{3/2}}, &  {\rm for}\quad x<x_{b}, \\
0, & {\rm for}\quad x>x_{b},
\end{array}\right.
\label{popovsaAA}
\end{equation}
where $\alpha$, $\beta$ are real constants.

The $Set$ $1$, $Set$ $3$ of analytic solutions to the Thomas-Fermi equation (\ref{eqless5g1})  belong to the more general following set
\begin{equation}
 \phi(x;a,b)= \left\{\begin{array}{ll} c_ {1}[\Phi(x;a,b)+c_ {2}], &  {\rm for}\quad x<x_{b}, \\
\frac{\alpha}{\beta+x}, & {\rm for}\quad x>x_{b},
\end{array}\right.
\label{popovsbH}
\end{equation}
where $\Phi(x,a,b)$, $c_ {1}$, $c_ {2}$ are given by
\begin{equation}
\left\{\begin{array}{ll} \Phi(x;a,b)=-\frac{x}{2(a-1)} {F}_{1;2}(1/2,a;3/2,-b^{2}x^{2}), \\
c_ {1}=\left(\lim _{x\rightarrow -k{\it Rc}}\Phi \left( x,a,b \right)+\lim _{x\rightarrow \infty }(\Phi \left( x,a,b \right) \right))^{-1}  & ,c_ {2}=\lim _{x\rightarrow \infty }\Phi \left( x,a,b \right) .
\end{array}\right.
\label{popovsaI}
\end{equation}
and $F_{1;2}$ is the Gauss hyper-geometric function.
For positive integer ($\geq 2$) or positive half-integer ($\geq 3/2$) values of $a$, $F_{1;2}$  can be written in terms of elementary functions (Table \ref{x}).
\begin{table}
\caption{Generalized exact solutions to $Set$ $1$ and $Set$ $3$ }
\label{x}
\begin{center}
\begin{tabular}{|c|c|}
	\hline
  $a$ &  $\Phi(x;a,b)$     \\
  \hline
  $3/2 $  & $- \sinh^{-1} (bx)$   \\
  $ 2$   & $-\frac{1}{2b}\arctan (bx)$ \\
  $  5/2 $  &  $-\frac{x}{\sqrt{3(1+b^{2}x^{2})}}$  \\
       $     3 $  &  $-\frac{x}{(8(1+b^{2}x^{2}))}-\frac{1}{8b}{\arctan (bx)}$  \\
         $  7/2 $ & $-\frac{x(3+2b^{2}x^{2})}{(15(1+b^{2}x^{2})^{3/2})}$    \\
         $   4  $ & $-\frac{x}{(24(1+b^{2}x^{2})^{2})} -\frac{x}{(24(1+b^{2}x^{2}))}-\frac{1}{16b}{\arctan (bx)}$   \\
          $ 9/2  $ & $-\frac{x(15+20b^{2}x^{2}8b^{4}x^{4})}{(105(1+b^{2}x^{2})^{5/2})}$   \\
           $ . $  & $.$   \\
           $ . $  & $.$   \\
          $  . $  & $.$   \\
    \hline
\end{tabular}
\end{center}
\end{table}

Also the $Set$ $2$  belongs to the more general set given by
\begin{equation}
 \phi(x;a,b)= \left\{\begin{array}{ll} c_ {1}[\Phi(x;a,b)+c_ {2}], &  {\rm for}\quad x<x_{b}, \\
\frac{\alpha}{\beta+x}, & {\rm for}\quad x>x_{b},
\end{array}\right.
\label{popovsbL}
\end{equation}
where 
\begin{equation}
 \left\{\begin{array}{ll} \Phi(x;a,b)=\int_{-kR_{c}}^{x} \frac{b}{2a}(1-\tanh(by)^{2})^{a} dy , \\
c_ {1}=\left(\lim _{x\rightarrow -k{\it Rc}}\Phi \left( x,a,b \right)+\lim _{x\rightarrow \infty }\Phi \left( x,a,b \right) \right)^{-1} , & c_ {2}=\lim _{x\rightarrow \infty }\Phi \left( x,a,b \right) .
\end{array}\right.
\label{popovsaM}
\end{equation}
For positive integer ($\geq 1$) or positive half-integer ($\geq 1/2$) values of $a$, $F_{1;2}$  can be written in terms of elementary functions (Table \ref{x1}).
\begin{table}
\caption{Generalized exact solutions to $Set$ $2$}
\label{x1}
\begin{center}
\begin{tabular}{|c|c|}
	\hline
$a$ &  $\Phi(x;a,b)$     \\
	\hline
1/2      & $- \sinh^{-1} (\tanh(bx))$   \\
1        & $-\frac{1}{2}\tanh(bx)-\frac{1}{4}\ln(\tanh(bx)-1)+\frac{1}{4}\ln(\tanh(bx)+1)$ \\
3/2      &  $-\frac{1}{6}\tanh(bx)\sqrt{1-\tanh(bx)^{2}}- \frac{1}{6}\sinh^{-1} (\tanh(bx))$  \\
2        &  $\frac{1}{12}\tanh(bx)^{3}-\frac{1}{4}\tanh(bx)-\frac{1}{8}\ln(\tanh(bx)-1)+\frac{1}{8}\ln(\tanh(bx)+1)$  \\
5/2      & $-\frac{3}{40}\tanh(bx)\sqrt{1-\tanh(bx)^{2}} -\frac{1}{20}\tanh(bx)(1-\tanh(bx)^{2})^{3/2}- \frac{3}{40}\sinh^{-1} (\tanh(bx))$    \\
.   & $.$   \\
.   & $.$   \\
.   & $.$   \\
\hline
\end{tabular}
\end{center}
\end{table}

These results, obtained by explicit analytic formulae, complement the numerical results presented in \cite{jorge}.


\begin{thebibliography}{99}

\bibitem{rrx200613}
R. Ruffini, M. Rotondo  and S. S. Xue,  {\it Int. J. Mod. Phys. D} {\bf 16}(1), 1 (2007).


\bibitem{prrx2009}
V. Popov, M. Rotondo, R.Ruffini  and S. S. Xue,  {\it astro-ph 0903.3727v1}.



\bibitem{migdal7614}   A. B. Migdal,  D. N. Voskresenskii   and  V. S. Popov, {\it JETP Letters} {\bf 24}(3), 186 (1976).


\bibitem{jorge} J.A. Rueda H.,B.Patricelli, M.Rotondo, R.Ruffini and S-S. Xue,  to appear in {\it Proceedings of the Third Stueckelberg Workshop } ( Cambridge Scientific Publisher, 2009).



\end{thebibliography}
\end{document}